\documentstyle[graphicx]{article}
\begin{document}
{\bf Supersoft X-ray emission from a white dwarf binary not powered by nuclear fusion}

Thomas J. Maccarone$^1$, Thomas J. Nelson$^2$, Peter J. Brown$^3$, Koji Mukai$^4$, Philip A. Charles$^5$, Andry Rajoelimanana$^6$, David A. H. Buckley $^7$,Jay Strader$^8$, Laura Chomiuk $^8$, Christopher T. Britt $^8$, , Saurabh W. Jha$^9$, Przemek Mr\'oz$^{10}$, Andrzej Udalski$^{10}$, Michal K. Szyma\'nski$^{10}$, Igor Soszy\'nski$^{10}$, Rados\l{}aw Poleski$^{10,11}$, Szymon Koz\l{}owski$^{10}$, Pawe\l Pietrukowicz$^{10}$, Jan Skowron$^{10}$, Krzysztof Ulaczyk$^{10,12}$

$^1$ Department of Physics and Astronomy, Texas Tech University, Box 41051, Lubbock, TX, 79409, USA\\
$^2$ Department of Physics and Astronomy, University of Pittsburgh, 100 Allen Hall, 3941 O'Hara Street, Pittsburgh PA, 15260, USA\\
$^3$ George P. and Cynthia Woods Mitchell Institute for Fundamental Physics \& Astronomy, Department of Physics and Astronomy, 4242 TAMU, College Station, TX 77843, USA\\
$^4$ CRESST, X-ray Astrophysics Laboratory, NASA Goddard Space Flight Center, Greenbelt MD, 20771, USA\\
$^5$ Physics and Astronomy, Faculty of Physical Sciences and Engineering, University of Southampton, Southampton SO17 1BJ, UK\\
$^6$ Department of Physics, University of the Free State, 205 Nelson Mandela Drive, Park West, Bloemfontein, 9301, Republic of South Africa\\
$^7$ South African Astronomical Observatory, PO Box 9, Observatory, 7935, South Africa\\
$^8$ Department of Physics and Astronomy, Michigan State University, East Lansing MI, 48823, USA\\
$^9$ Department of Physics \& Astronomy, Rutgers, the State University of New Jersey, 136 Felinghuysen Road, Piscataway NJ 08854, USA\\ $^{10}$ Warsaw University Observatory, Al. Ujazdowskie 4, 00-478 Warszawa,
Poland\\
$^{11}$ Department of Astronomy, Ohio State University, 140 W. 18th Ave.,
Columbus, OH 43210, USA\\
$^{12}$ Department of Physics, University of Warwick, Gibbet Hill Road,
Coventry, CV4 7AL, UK\\

{\bf Supersoft X-ray sources are stellar objects which emit X-rays
  with temperatures of about 1 million Kelvin and luminosities well in
  excess of what can be produced by stellar coronae.  It has generally
  been presumed that the objects in this class are binary star systems
  in which mass transfer leads to nuclear fusion on the surface of a
  white dwarf.$^1$ Classical novae, the runaway fusion events on the
  surfaces of white dwarfs, generally have supersoft phases, and it is
  often stated that the bright steady supersoft X-ray sources seen
  from white dwarfs accreting mass at a high rate are undergoing
  steady nuclear fusion.$^1$ In this letter, we report the discovery
  of a transient supersoft source in the Small Magellanic Cloud
  without any signature of nuclear fusion having taken place. This
  discovery indicates that the X-ray emission probably comes from a
  “spreading layer”$^2$ - a belt on the surface of the white dwarf
  near the inner edge of the accretion disk in which a large fraction
  of the total accretion energy is emitted - and (albeit more
  tentatively) that the accreting white dwarf is relatively massive.
  We thus establish that the presence of a supersoft source cannot
  always be used as a tracer of nuclear fusion, in contradiction with
  decades-old consensus about the nature of supersoft emission.}

ASASSN-16oh was discovered by the All-Sky Automated Survey for
Supernovae as a V=16.9 transient, located in the field of the Small
Magellanic Cloud at right ascension of 1:57:43.64 and declination of
-73:37:32.5 (J2000), on December 2, 2016.  It was quickly determined to show
narrow optical emission lines (unresolved with a resolution of about
300 km/sec), at a velocity consistent with the SMC’s velocity and
hence to be an unusual transient object$^3$. Observations with the
Neil Gehrels Swift Observatory, using its X-ray
telescope taken on December 15, 2016 then showed a supersoft X-ray
source, with a temperature of about 900,000 K, and a luminosity of
about $1.0\times 10^{37}$ erg/sec, assuming the source is located in
the SMC$^4$.  The pre-outburst emission taken from the OGLE-IV
monitoring program showed that the source has been an irregular
variable for several years, with quiescent fluxes of about I=20.3, and
V=21.1$^5$ (see also Figure 4 in the supplementary information).

On December 28, 2016, the Chandra X-ray Observatory observed this
object for 50 kiloseconds with the Low Energy Transmission Grating.
The observation revealed a purely continuum spectrum longward of 15
~\AA, with little evidence of superimposed emission lines.  The data
can be well modelled by either a blackbody with a temperature of
905,000$\pm$50000 K and a luminosity of $6.7^{+1.5}_{1.2}\times
10^{36}$ erg/s at the distance of the SMC$^6$, or a non-local thermal
equilibrium atmosphere model$^7$ with abundances similar to the Milky
Way halo which results in a similar quality fit to the blackbody
(C-statistic/dof = 2890/2796) and constrains the atmospheric
temperature to 750000 (+30000, -15000) K absorbed by
$2.0\pm0.5\times10^{20}$ cm$^{-2}$.  A post-eruption nova model$^7$ gives
a slightly worse fit, and constrains only a lower limit to the
temperature of 1.05 MK.

The supersoft X-ray emission cannot come from a nuclear fusion
episode.  There is no sharp variation in the optical light curve from
the source indicating that the source “turned on” as a fusion source,
and the peak absolute magnitude of the event is about 5 magnitudes
fainter than those typical of classical novae { (and 3.9 magnitudes
  less luminous than the peak of the M31 system which has recurrent
  nova eruptions on sub-year timescales$^{8}$)}, and the rise to
maximum brightness much longer than the typical nova timescale of a
few days.  { Most importantly,} optical spectroscopy obtained during
the supersoft phase continued to show narrow emission lines from
hydrogen and helium, in agreement with a disk origin for the line
emission, but contrary to expectations for nova ejecta { -- thus,
  even a very short burst of fusion, quenched by a strong wind can be
  excluded}.

{ The emission must, nonetheless, be coming from a small fraction of
  the surface area of the white dwarf.  Three possibilities exist:
  magnetic collimation of the flow into small hot spots (which is
  unlikely due to the high magnetic field required and the lack of
  evidence for rotational modulation of the emission), a classical
  boundary layer$^{9}$ or a spreading layer$^{2,10,11}$.  The latter
  two mechanisms are different models for how the rotational energy in
  the flow is dissipated as the Keplerian inner disk joins to the more
  slowly rotating white dwarf.  The standard boundary layer predicts
  ${T}\propto{\dot{M}^{1/4}}$, and hence should show a factor of about
  3 variation in temperature for the source SS Cyg when it varies by a
  factor of 100 in luminosity, but no evidence for a temperature
  change was seen.$^{12}$, so we focus on the spreading layer
  interpretation, but note that there are still unsettled questions in
  the details of the structures of the inner accretion flows, so that
  the parameter values inferred have substantial uncertainties due to
  model dependencies.}

The temperature of the spreading layer is calculated to be:$^{2}$
\begin{equation}
T = 2\times10^5 {\rm K} \alpha_{disk}^{-3/40} \alpha_3^{1/8} \dot{M}_{18}^{9/80} M_1^{13/32} R_9^{-23/32}
\end{equation}
where $\alpha_{disk}$ is the dimensionless viscosity parameter of the
accretion disk which is typically taken to be 0.1, $\alpha_3$is the
dimensionless viscosity parameter of the spreading layer in units of
$10^{-3}$, $\dot{M}_{18}$ is the mass accretion rate in units of
$10^{18}$ g/sec, $M_1$ is the mass of the white dwarf in solar masses,
and $R_9$ is the radius of the white dwarf in units of $10^9$ cm.
There are stronger dependances on the mass of the white dwarf than on
the accretion rate, and these are also the two parameters which have
the greatest range of reasonable possible values.  The observed
temperature of 900,000 K can be achieved for a mass transfer rate of
$2\times10^{19}$ g/sec, a white dwarf mass of 1.3 $M_\odot$, and a
white dwarf radius of $3\times10^8$ cm with the other values left at
their fiducial values.  Taking the atmosphere model temperature,
rather than the blackbody model will drop the white dwarf mass to
about 1.2 $M_\odot$ and raise its radius to about $4\times10^8$ cm.
{ Unless the viscosity values are far from their fiducial values,
  in the context of the spreading layer model} we require both a white
dwarf well in excess of a solar mass, and on which the bulk of the
emission is coming from about 10\% of the surface area, because we
cannot otherwise explain the combination of the high temperature and
the only moderately large luminosity from the source.  { We further
  note some caution in using the numbers here, since they are at the
  extremes of the model calculation, and some processes like advective
  heating of the white dwarf due to photon trapping may be more
  important for such high accretion rates than they are at more
  typical accretion rates.  It can thus be robustly stated that the
  emission region must be substantially smaller than the surface area
  of the white dwarf, and stated only in a model-dependent manner that
  the white dwarf is likely to be toward the upper end of the mass
  range for white dwarfs.}

\begin{figure}
\includegraphics[width=5.5 in]{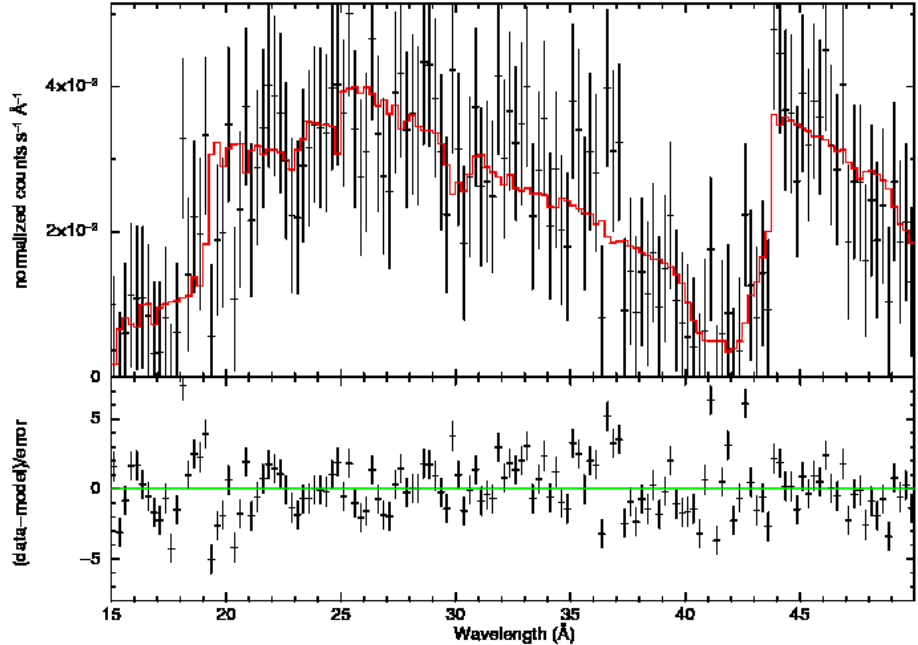}
\caption{The best fitting Chandra spectrum of ASASSN~16-oh using a
  non-LTE model for the supersoft source atmosphere.$^7$ One can see
  that there are no strong, structured residuals. The red line in the
  top panel and the green line in the bottom panel are both the model
  fit to the data.  Any real spectral line from a nova would be several
  wavelength bins across, and it is clear from the spectrum that there
  are no such lines.  A 905,000$\pm$50000 K blackbody model absorbed
  by $3.4\pm0.5\times10^{20}$ cm$^{-2}$ provides an acceptable fit to
  the data (C-statistic/dof = 2880/2796). The implied luminosity is
  $6.7\pm^{1.5}_{1.2}\times 10^{36}$ erg/s at the distance of the
  SMC$^6$. Although no absorption lines are apparent in the residuals
  of the blackbody fit at the 2-3 sigma level, we also tried two
  Non-Local Thermal Equilibrium (NLTE) atmosphere models appropriate
  for hot, compact stars$^7$, to assess the possible range of
  temperatures consistent with the observed spectrum. A model
  developed for post-eruption novae gives a slightly worse fit
  (C-statistic/dof = 2903/2796) and only a lower limit to the
  temperature of 1.05 MK, while a model with abundances similar to the
  Milky Way halo results in a similar quality fit to the blackbody
  (C-statistic/dof = 2890/2796) and constrains the atmospheric
  temperature to 750000 (+30000, -15000) K absorbed by 2.0$\pm
  0.5\times 10^{20}$ cm$^{-2}$.The apparent strong features in the
  spectrum are at the wavelengths of the carbon and oxygen edges.
  Error bars are 1-sigma.}
\end{figure}

Observation of a transient, then, with { this combination of
  luminosity and temperature} requires a larger peak accretion rate
than in typical cataclysmic variable outbursts.$^{13}$ Tentative
suggestions have been made that other systems have previously shown
this phenomenon$^{13}$ (e.g. SS Cyg, in which the spreading layer may
have been responsible for its extreme ultraviolet emission), but
without considerable monitoring, it is not possible to establish this
clearly.  The finding here, of a temperature which cannot otherwise be
reached without nuclear fusion { demonstrates that multiple
  mechanisms exist for producing supersoft sources.}

The high accretion rate from ASASSN-16oh can be explained by the fact
that, having a highly evolved donor star, its orbital period must be
long and its accretion disk must be large, so that the disk has a
large mass reservoir when its outburst is triggered. This combination
of factors can lead to both a high quiescent luminosity with a large
fraction of the quiescent light from the accretion flow and a very
high peak $\dot{m}$ during dwarf nova outbursts.

Several lines of argument indicate that this system must be at an
orbital period of a few days. Theoretical predictions of the relation
between the peak luminosity in the optical and the orbital period for
dwarf novae, which have been established to be in good agreement with
the empirical data, yield an expected orbital period of about 4 days,
while there is a hint of an orbital period of about 5.6 days seen in
optical spectroscopy, and the optical through ultraviolet spectral
energy distribution from Swift also suggests an orbital period of
about 5 days.  Additionally, two other cataclysmic variables with
orbital periods greater than one day show similar optical
phenomenology, of very slow rise times to their outbursts, and very
bright peaks$^{14,15,16}$ and may be indicative of these longer period
systems undergoing “inside-out” outbursts where the ionization
instability is triggered in the inner part of the accretion disk
rather than at the outermost part.$^{17}${ It would be very
  valuable to look for soft X-ray emission associated with the next
  outbursts of these systems.}

In addition to the discovery’s implications for the physics of
accretion onto white dwarfs, this result has additional important
implications for understanding the source populations of supersoft
sources.  It is often assumed that all supersoft X-ray sources are
either classical novae or accreting white dwarfs undergoing steady
nuclear fusion.  Our finding of a transient supersoft source which is
clearly not a classical nova, due to its lack of broad optical
emission lines, shows that there exists an important alternative
mechanism for producing supersoft X-ray emission.  The results may
help explain the previously mysterious subluminous supersoft sources
seen in nearby galaxies.$^{18}$

{\bf References}\\
1.  Kahabka P \& van den Heuvel E.P.J., Luminous Supersoft X-ray Sources, {\it Annu. Rev. Astron. Astrophys.}, {\bf 35}, 69-100 (1997)\\
2.  Piro A.L., \& Bildsten L., Spreading of Accreted Material on White Dwarfs, {\it Astrophys. J.}, 610, 977-990 (2004)\\
3.  Jha, S.W., et al., ASASSN-16oh: An Unusual Transient in the Vicinity of the SMC,  {\it Astronom. Telegram}, 9859 (2016)\\
4.  Maccarone T.J., Brown P., Mukai K., Swift observatons of ASASSN-16oh,  {\it Astronom. Telegram}, 9866 (2016)\\
5.  Mroz P., et al., OGLE-IV Observations of ASASSN-16oh, {\it Astron. Telegram}, 9867 (2016)\\
6.  Graczyk D. et al, The Araucaria Project. The Distance to the Small Magellanic Cloud from Late-type Eclipsing Binaries, {\it Astrophys. J.}, {\bf 780}, 59-71 (2014)\\
7.  Rauch T., Orio M., Gonzales-Riestra R., Nelson T., Still M., Werner K., Wilms J., Non-local Thermal Equilibrium Model Atmospheres for the Hottest White Dwarfs: Spectral Analysis of the Compact Component in Nova V4743 Sgr, {\it Astron. J.}, {\bf 717}, 363-371 (2010)\\
8. Darnley M.J., et al., M31N 2008-12a -- The Remarkable Recurrent Nova in M31: Panchromatic Observations of the 2015 Eruption. {\it Astrophys. J.}, {\bf 833}, 149, (2016)\\
9. Pringle J.E.,  Soft X-ray emission from dwarf novae,{\it Mon. Not. R. Astron. Soc.} , {\bf 178}, 195 (1977)\\
10. Inogamov N.A., Sunyaev R.A., Spread of matter over a neutron-star surface during disk accretion,  {\it Astron. Let.}, {\bf 25}, 269 (1999)\\
11. Kippenhahn R., Thomas H.-C., Accretion belts on white dwarfs, {\it Astron. \& Astrop.}, {\bf 63},265, (1978)\\
12. Mauche C.W., Raymond J.C., Mattei J.A., EUVE Observations of the Anomalous 1993 August Outburst of SS Cygni, {\it Astrophys. J.}, {\bf 446}, 842 (1995)\\
13.  Warner B., Absolute magnitudes of cataclysmic variables,{\it Mon. Not. R. Astron. Soc.}, {\bf 227}, 23-73 (1987)\\
14.  Salazar I.V., LeBleu A., Schaefer B.E., Landolt A.U., Dvorak S., Accurate pre- and post-eruption orbital periods for the dwarf/classical nova V1017 Sgr, {\it Mon. Not. R. Astron. Soc.} , {\bf 469}, 4116-4132 (2017)\\
15.  Shears J. \& Poyner G., The 2009 outburst of V630 Cassiopeiae, {\it J. British Astron. Association}, {\bf 120}, 169 (2010)\\
16.  Orosz J., Thorstensen J.R., Honeycutt R.K.,The long-period orbit of the dwarf nova V630 Cassiopeiae, {\it Mon. Not. R. Astron. Soc.} , {\bf 326}, 1134-1140 (2001)\\
17.  Smak J., Outbursts of dwarf novae, {\it Proc. Astron. Soc. Pacific}, {\bf 96}, 5-18 (1984)\\
18. Di Stefano R., Kong, A.K.H., The Discovery of Quasi-soft and Supersoft Sources in External Galaxies, {\it Astrophys. J.}, {\bf 609}, 710 (2004)\\

{\bf Acknowledgments:} We thank Jeno Sokoloski, Denija Crnojevi\'c and
Chris Sneden for useful discussions.  We thank Belinda Wilkes and the
CXC staff for approving and executing a director’s discretionary time
observation.  The OGLE project has received funding from the National
Science Center, Poland, grant MAESTRO 2014/14/A/ST9/00121 to A.U.  PAC
acknowledges support of the Leverhulme Trust.  Some of these
observations were done with the Southern African Large Telescope under
program 2016-2-LSP-001. DB acknowledges support of the National
Research Foundation.

{\bf Data availability statement} The data from Chandra and Swift are
available from the NASA HEASARC repository.  The first two SALT
spectra are available from
https://wis-tns.weizmann.ac.il/object/2016irh while the additional
SALT spectra are available by contacting Andry Rajoelimanana.  The
remaining SALT spectra are available from
https://cloudcape.saao.ac.za/index.php/s/qeodvvMP1TLIy4H.  The
remaining other data that support the plots within this paper and the
other findings in this study are available from the corresponding
author upon reasonable request.

{\bf Competing interests} The authors declare that they have no
competing financial interests.

{\bf Contributions}

Maccarone wrote one of the proposals for Swift observations, reduced
and analyzed the Swift X-ray data, modelled the ultraviolet data,
wrote the Chandra proposal, contributed heavily to the interpretation
and wrote most of the paper text.  Nelson analyzed the Chandra data,
and contributed heavily to the interpretation.  Brown wrote one of the
Swift proposals, and reduced the Swift UVOT data and contributed to
the interpretation.  Mukai first proposed the spreading layer
hypothesis and contributed heavily to the interpretation beyond that.
Charles, Strader, Rajoelimanana, Britt, Chomiuk, Buckley, Jha, Mr\'oz,
Udalski, Szyma\'nski, Soszy\'nski, Poleski, Koz\l{}owski, Pietrukowicz,
Skowron and Ulaczyk provided supporting optical data.  All authors
reviewed the paper draft.

{\bf Methods}

The Swift UVOT data were extracted following the standard procedure$^{19}$.
Swift observed the source near peak brightness on MJD 57737.44 and
57738.83.

We fit the Swift ultraviolet and optical data from the first two
observations, for which there were 6 filters of data taken, using a
power law model.  For the observation made on MJD 57737.440, the
spectrum is consistent with a power law in $F_\lambda$ which is
proportional to $\lambda^{-2.33\pm0.14}$, assuming a reddening law
typical of the SMC$^{20}$, and that the value of $N_H$ is that from the
X-ray data, of $2\times10^{20}$ cm$^{-2}$.  Even assuming no
reddening, the spectral index changes only from -2.33 to -2.28, so,
while the SMC reddening law is variable, and not fully characterized,
the effects of uncertainty in reddening are smaller than the
statistical uncertainty of the measurements.  A nearly simultaneous
OGLE data point in the I-band is consistent with the extrapolation of
the power law.

The observation made the next day shows essentially the same results.
The estimated spectral index is consistent with the law, $F_\nu
\propto \nu^{1/3}$ expected for a Shakura-Sunyaev accretion disk in
intermediate wavelength regime (equivalent to $F_\lambda \propto
\lambda^{-2.33}$), at wavelengths longer than the wavelength of the
Wien peak of the inner part of the accretion flow, and wavelengths
shorter than the wavelength of the Wien peak of the outer part of the
accretion disk.

We can then take the mass transfer rate inferred from the X-ray data,
and use the Shakura-Sunyaev disk’s relation between temperature and
disk radius:
\begin{equation}
T = 4.1\times10^4 {\rm K} \dot{M}_{16} R_9^{-3/4}
\end{equation}
and we find that if the outer temperature is about 6000 K, consistent
with seeing an outburst in progress, and with seeing the emission
showing no clear break even in the V band, then the outer disk radius
should be about $1.4\times10^{11}$ cm.  If the outer disk radius is
the circularization radius of the accretion disk, then the orbital
period should be about 5 days, assuming a white dwarf of 1.3 $M_\odot$
and a donor of about 0.7 $M_\odot$.  If, however, the accretion disk
spreads outwards to the Roche lobe radius of the white dwarf during an
outburst, then the orbital period may be closer to 1 day than to 5
days.

\begin{figure}
\includegraphics[angle=-90,width=6 in]{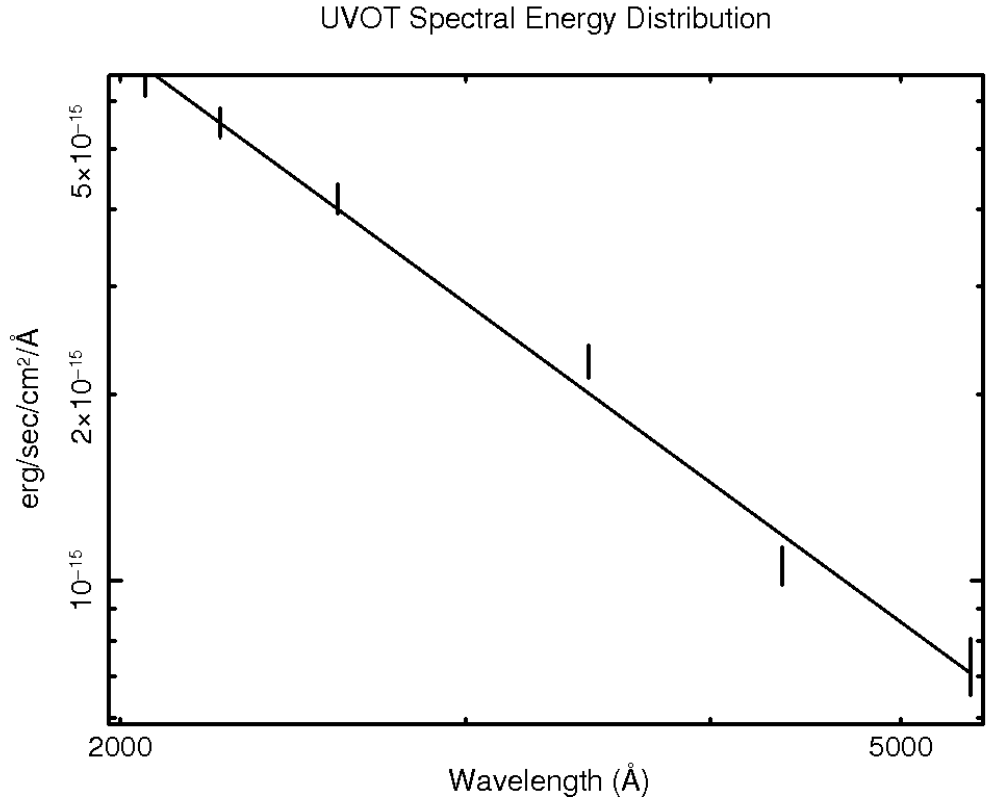}
\caption{The ultraviolet spectra energy distribution of the new SMC
  transient, with the flux density in erg/sec/cm$^2$/\AA  plotted
  versus the wavelength in Angstroms.  The fit presented is for a
  power law with a spectral index of -2.33 for $F_\lambda$, which is
  consistent with the expectations for a thermal accretion disk with
  radiated power in each annulus equal to the gravitational energy
  release in the annulus. Error bars are 1-sigma.}
\end{figure}

{\it Swift X-ray spectroscopy details}

CCD-resolution X-ray spectra were obtained by Swift on 15 December
2016, for 993 seconds of on-source time.  The data were extracted
using xselect, with a 113 arcsec region around the source, and fitted
using the Cash statistic.  Taking advantage of the information about
the foreground absorption from the Chandra data, we fix the foreground
absorption to a column density of $2\times10^{20}$ cm$^{-2}$.  Under
these considerations, if we fit the spectrum from 0.5-1.0 keV, we find
a blackbody temperature of 0.068$\pm$0.01 keV, with a c-statistic of
20.7 for 26 degrees of freedom, and if we include also the range from
0.4-0.5 keV, the fitted temperature increases to 0.08$\pm$0.01 keV
with a c-statistic of 33.3 for 37 degrees of freedom.  Including the
lower energy channels gives an unabsorbed bolometric flux of
$1.5\times10^{-11}$ erg/sec/cm$^2$, while excluding them gives
$3.6\times10^{-11}$ erg/sec/cm$^2$.  Given that the data are probably
not a true blackbody, and that the fit is based on 176 total photons,
we can determine clearly that the spectrum is already that of a
supersoft source on 15 December, and that the X-ray luminosity is
about $10^{37}$ erg/sec, but cannot make significantly more detailed
claims from the Swift data.

{\it Chandra X-ray Spectroscopy details}

The Chandra LETG data were analyzed using CIAO version 4.8 and the
calibration database version 4.7.2.  New data products were extracted
using the chandra\_repro script, which creates screened event and
spectrum files.  In order to increase the signal to noise for spectral
modeling, we summed the $\pm$1st order spectra from the LETG using the
combine\_grating\_spectra script, which creates summed ancillary
response files and updated response matrix files.

Models were explored in XSPEC v.12.9.0n, and fit to the summed spectra
with no additional binning.  There is very little signal below 15 \AA
in the resulting spectrum, so we fit models in the wavelength range
15-50 \AA.  Best fit parameters were determined by minimizing the Cash
statistic.

{\it Rough orbital period estimation}

A few approaches can be used to establish that the system is very
likely to have an orbital period of several days, but continued
spectroscopic montioring will be needed to make a definitive
measurement of its orbital period.  These approaches are examination
of the outbursting and quiescent brightnesses of the source;
examination of the physical widths of the emission lines seen in
outburst, and examination of the range of radial velocities seen in
outburst.

Taking the Swift photometry near the peak of the outburst, with
V=16.8, applying a de-reddening correction of 0.03 magnitudes, based
on the $N_H$ value from the X-ray fitting, we get $M_{V,peak}$of -2.7 for a distance modulus of 18.95 to the SMC.$^{6}$  We can take the expression:$^{21}$
\begin{equation}
\dot{M}=1.1\times10^{-8} M_\odot{\rm/yr} \left(\frac{\alpha_{hot}}{0.1}\right)^{1.14}\left(\frac{\alpha_{cold}}{0.02}\right)^{-1.23}\left(\frac{r_{outer}}{4\times10^{10} {\rm cm}}\right)^{2.57} \left(\frac{f}{0.4}\right)^{1.43}
\end{equation}
where $\alpha_{hot}$is the viscosity parameter for the hot state of the disk, $\alpha_{cold}$ is the viscosity parameter for the cold state of the disk, $r_{outer}$ is the outer radius of the disk, which should be in some range of at least the circularization radius but no more than the tidal radius for the accretion disk, and $f$ is the fraction of the disk’s energy reservoir that is dumped in the transient outburst.  That expression matches well to the data$^{13}$ for the systems with orbital periods less than 10 hours.   

If we then take all the values except $r_{outer}$ at their fiducial
values, and taking $\dot{M}$ of $3\times10^{-7} M_\odot$/yr at peak
based on the X-ray spectra, we then find $r_{outer}$ of about
$1.4\times10^{11}$ cm, in good agreement with the ultraviolet and
optical SED.  We then find about 90 hours to be the likely orbital
period assuming a mass ratio of 0.5 for the donor star to the accretor
if the outer disk radius at the time of the triggering of the outburst
is the circularization radius.  This period is between those of the
two similar bright, long period cataclysmic variables with
“triangular” shaped optical outbursts , V630 Cas and V1017 Sgr, and
the source then also has a peak luminosity between the peak
luminosities of those two sources.$^{14,15}$

The pre-outburst fluxes of the star are I=20.3 and V=21.1$^{1}$.  The
pre-outburst light curve from OGLE shows $\approx$0.5 magnitudes of
variability, so that the colors estimated from non-simultaneous data
are not likely to be reliable.  The variability also indicates that
there is likely to be substantial accretion light contributed even in
quiescence.

We can additionally determine whether it is feasible to have a donor
star at approximately this brightness filling a Roche lobe in a binary
with a period of about 90 hours.  If we assume a donor mass of 0.7
$M_\odot$, then to fill this Roche lobe, we expect a radius of
3.9$R_\odot$. It is possible that a lower mass donor could be
similarly bloated, if it has followed an evolutionary pathway similar
to the companions to some of the millisecond pulsars$^{6,7}$, or that a
somewhat larger donor mass could be present without leading to
dynamically unstable mass transfer.  If a lower (or higher) mass is
taken, then the expected radius will also be lower (or higher).

If we then assume the star to have the colors of a late K dwarf, but
to be about 40 times brighter due to the larger radius, we should
expect it to be at an absolute magnitude of 5 in V, and 3.2 in I and
1.3 in K.  The distance modulus of 18.95 for the SMC then implies that
the star should be about V=24.1, I=22.3, K=20.4, fainter than
catalogued limits.  The object is detected by Spitzer in the SAGE
project,$^{22}$ although with $m_{3.6}$=18.43$\pm$0.17 and
$m_{4.5}$=17.31$\pm$0.16 (where these are the Vega magnitudes in the
band with center wavelength of the subscript in microns) giving a
color of 1.12$\pm$0.23.  This combination of brightness and color
likely indicates that some local dust emission is taking place from
the source, perhaps from a old nova shell, since at the 2 sigma
level, its Spitzer color is inconsistent with the colors of late L
dwarfs.

This would suggest that even in quiescence, most of the pre-outburst
optical light is from the accretion disk, making this donor type
plausible and still allowing for the observed substantial aperiodic
variability outside the main outburst peak.  Assuming that the light
from the binary includes a moderate sized accretion component making
the binary bluer and brighter than the donor star alone, a reasonable
agreement can be found, but that it may be difficult to identify
absorption lines for spectroscopy, or ellipsoidal modulations in the
optical band unless the donor star is slightly hotter than this.
Follow-up in the infrared may thus be the most promising approach to
obtain radial velocities from absorption lines in quiescence, despite
the source's infrared faintness.

{\it X-ray timing analysis} 

We produced a light curve from the zero order grating component (we
ignored the dispersed component to avoid background contamination).
The data were extracted with 1 second time resolution using {\textsc
  dmextract}, and then Fourier transformed, making use of the first
49152 seconds of data to make 6 segments of 8192 seconds each,
excluding a few percent of the data for convenience.  No peak in the
periodogram was statistically significant, and the $4-\sigma$ upper
limit on the rms fractional variability is 7\% over a range of periods
from 2 seconds to 8192 sec.

The lack of an X-ray periodicity is a moderately strong argument
against the possibility that the soft X-rays come from optically thick
polar accretion -- the known high accretion rate intermediate polar
source .  Additionally, at $\dot{m}$ = $2\times10^{19}$ g/sec, the
magnetic field would need to be about $10^6$ G in order to allow the
Alfven radius to be outside the white dwarf by a factor of 2.

{\it Optical line properties}

Optical spectra were obtained with the Robert Stobie Spectrograph
(RSS) on the 10 m Southern African Large Telescope (SALT) in long-slit
mode. We first observed with the PG0300 grating (on 2016-Dec-07) and
the PG0900 grating (on 2016-Dec-13), finding unresolved Balmer and
helium emission lines in the early work.3 We then began doing higher
resolution spectroscopy using either the grating PG2300 at an angle of
30.5 degrees or the grating PG3000 at an angle of 44.0 degrees with a
slit width of 1.5”, giving a resolving power of $2000<R<3800$. The
data were pre-reduced using the PYSALT package which includes bias
subtraction, gain and cross-talk corrections and
mosaicking. Wavelength calibrations and spectra extractions were
performed with the IRAF long-slit package.

The number and time-sampling of the optical spectra is currently not
sufficient to make a definitive detection of an orbital period.  We
can, however, make a good estimate of the radial velocity amplitude of
the binary from the range of radial velocities seen.  The velocities
span a range of about 75 km/sec.  Additionally, the observed lines are
fairly narrow, with the average full width at half-maximum of the
lines being 164 km/sec.  Both of these findings require either that
the orbital period of the binary be of order days, or that the binary
is exceptionally face-on.  If we assume, e.g. that the emission lines
come from the accretion disk and are a good tracer of the orbital
motions of the accretor, that the donor mass is 0.7 solar masses, and
that the accretor mass is 1.3 solar masses, then an inclination of 12
degrees is required for a 1 day orbital period and an inclination
angle of 27 degrees is required for a 10 day orbital period.  Lower
inclination angles would be required for a lower mass donor star.

The lack of broad optical lines also provides additional evidence that
the supersoft source is being produced without nuclear fusion.  The
emission lines presumably are produced over a range of radii in the
accretion disk, with the bulk of the flux coming from the outer part
of the disk where the emitting area is largest.  For disk lines, for a
given line width or radial velocity amplitude there is a degeneracy
between the actual physical velocity and the inclination angle of the
system.  There are well-calibrated relations between the velocity
width of emission lines in quiescent X-ray binaries and CVs$^{23}$, but
these relations are not well calibrated during outbursts where the
lines can be expected to become narrower as the inner regions of the
disks become fully ionized.  We find that half the range of radial
velocity variations, about 38 km/sec is about 0.23 times the full
width half-maximum of the emission lines.  This is the same as the
ratio found in X-ray binaries in quiescence$^{23}$ for the semi-amplitude
of the donor star’s radial velocity curve.  Since here the emission
lines are likely to be tracing the motion of the heavier component,
the accretor, this is suggestive of the aforementioned narrowing of
the emission lines in outburst due to full ionization of the inner
regions of the disk.

Classical novae are quasi-spherical explosions, so that regardless of
inclination angle, they should produce broad lines.  The spectra in
figure 3 illustrate that the emission lines observed from ASASSN-16oh
are narrow over the range of dates between the first Swift observation
showing that the source is supersoft and the Chandra observation that
better characterizes the supersoft spectrum.  While a transient
outburst might trigger a long-lived episode of steady nuclear fusion,
it is highly unlikely that it could do so without first triggering a
nova eruption of the fuel that piled up during the non-fusing phase of
accretion.

\begin{figure}
\includegraphics[width = 5 in]{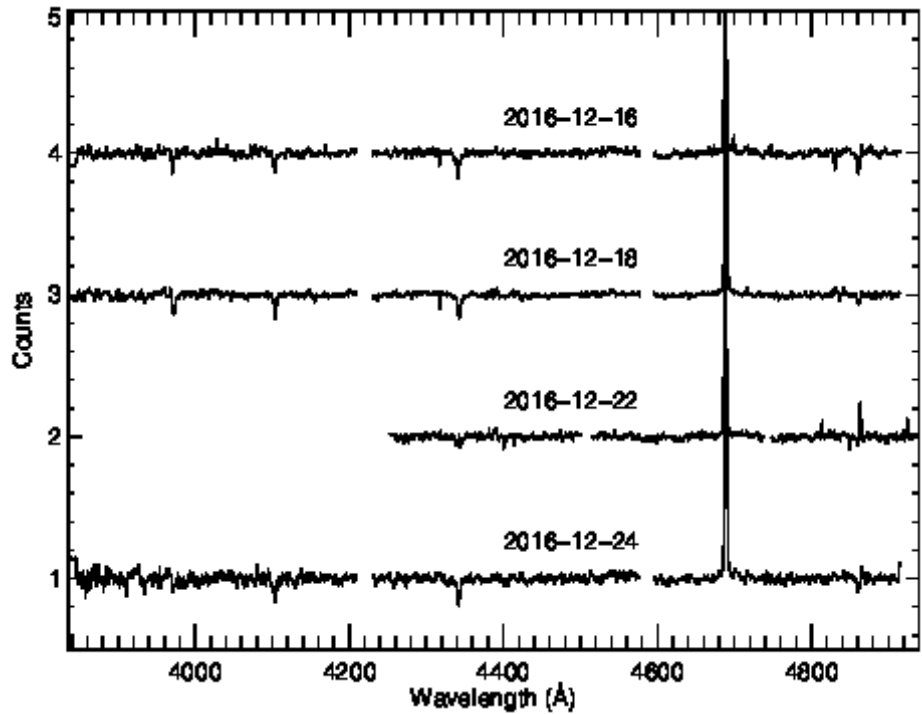}
\caption{SALT optical spectra taken just after the optical maximum (14th December 2016) with grating PG2300 and PG3000. The gaps in the spectra are due to CCD gaps. The average FWHM of the HeII lines is 3.04 \AA, or 2.57 \AA, equivalent to 164 km/sec, after correcting for instrumental broadening.  The equivalent width is 5.24 Angstrom.}
\end{figure} 	 	

{\it The maximum accretion rate that will avoid fusion}

Next, we note that the accretion rate we expect for this object is
about $3\times10^{-7} M_\odot$/yr, while state-of-the-art theory
calculations suggest that $3.5\times10^{-7} M_\odot$/yr$^{24}$ is the
threshold accretion rate above which steady nuclear fusion will take
place for an accretor which is $1.3 M_\odot$ white dwarf.
{ Importantly, however, the system is not accreting steadily at that
rate.  Only if it accumulates a large enough envelope at such a high
mass transfer rate will there be stable fusion.  The full-width,
half-maximum of the peak is only about 50 days, meaning that only
about $4\times10^{-8} M_\odot$ is accumulated during a period of time
with $\dot{M}$ close to the critical level.  This is too low an
envelope mass to allow steady burning, even on the most massive white
dwarfs.$^{25}$}

{\it The long term optical light curve from the source}

A long term optical light curve for this object has been obtained by
the Optical Gravitational Lensing Experiment$^{26}$.  It shows a few salient
features, in addition to providing good sampling of the bright
outburst.  The I-band light curve is plotted in Figure 4.  A smaller
number of V-band data points have also been obtained by OGLE, and
these indicate that the V-I colors are relatively steady at about
V-I=0.8 until the major outburst begins.  This color is similar to the
color of a late G star (i.e about 5300 K).  If the light in the V and
I bands is dominated by accretion disk light, then this temperature is
consistent with the expectations for a disk that will be susceptible
to the dwarf nova instability, but will still be bright enough to
dominate the optical flux relative to a red evolved donor star.  The
large level of variability before the outburst does suggest that the
donor star is not the dominant contributor to the quiescent optical
flux.  This finding is similar to what has been seen in some
long-orbital period X-ray binaries, even in “quiescence”.$^{27}$
Additionally, it is clear from the light curve that there is no sharp
change in the optical flux suggestive of a sudden triggering of
nuclear fusion.

\begin{figure}
\includegraphics[width=5 in]{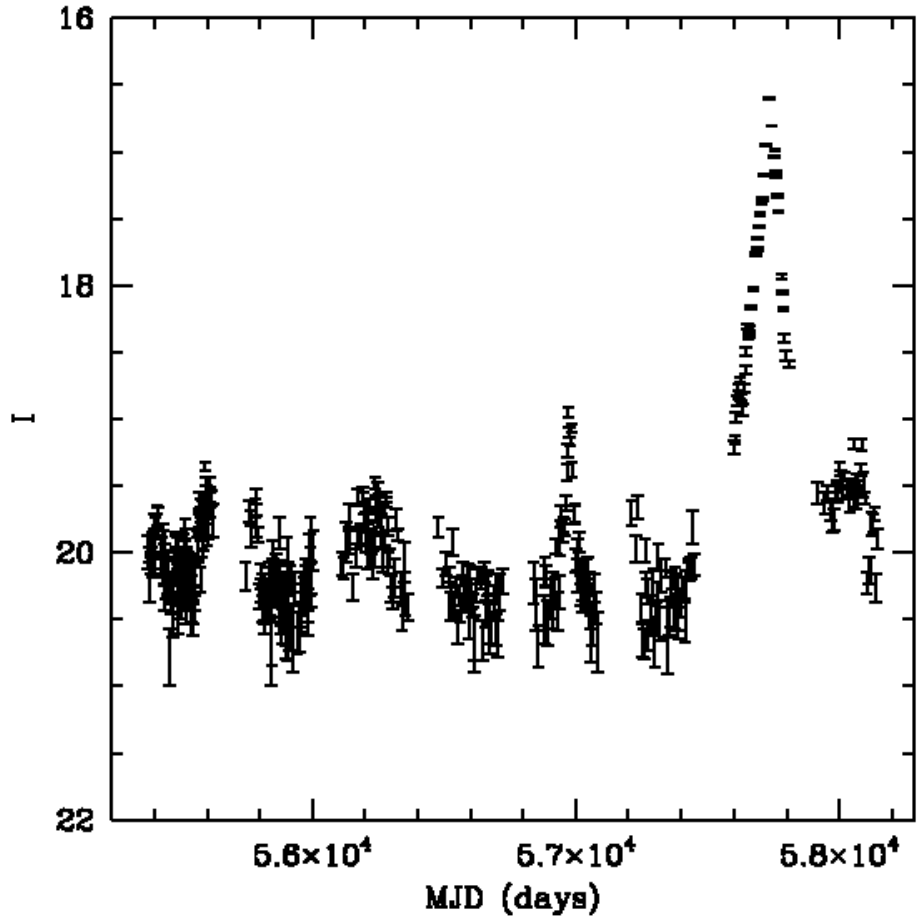}
\caption{The OGLE I band light curve for ASASSN-16oh.  The light curve shows that there are no sharp spikes in the optical time series for the source that might indicate a start of nuclear fusion and that there is substantial quiescent variability for the source.  The OGLE colors are generally about V-I of 0.8, except for the three data points taken during the large outburst, which are significantly bluer (0.36$\pm$0.02 on MJD 57657, 0.34$\pm$0.02 on MJD 57681 and 0.16$\pm$0.01 on MJD 57744). Error bars are 1-sigma.}
\end{figure}

{\bf References}

19.  Brown P.J. et al.,Ultraviolet Light Curves of Supernovae with the Swift Ultraviolet/Optical Telescope, {\it Astron. J.}, {\bf 137}, 4517-4525 (2009)\\
20. Gordon K., Clayton G.C., Misselt K.A., Landolt A.U., Wolff M.J., 2003, {\it Astrophys. J.}, {\bf 594}, 279-293\\
21. Cannizzo J.K., The Accretion Disk Limit Cycle Model: Toward an Understanding of the Long-Term Behavior of SS Cygni, {\it Astrophys. J}, {\bf 419}, 318-336 (1993)\\
22.  Gordon K.D., et al., {\it Astron. J.}, {\bf 142}, 102-116 (2011)\\
23. Casares J., A FWHM-K2 Correlation in Black Hole Transients, {\it Astrophys. J.}, {\bf 808}, 80-90 (2015)\\
24. Wolf W.M., Bildsten L., Brooks J., Paxton B., Hydrogen Burning on Accreting White Dwarfs: Stability, Recurrent Novae, and the Post-nova Supersoft Phase, {\it Astrophys. J.}, {\bf 777}, 136-150 (2013)\\
25. Kato M., Saio H., Hachisu I., Nomoto K., Shortest Recurrence Periods of Novae, {\it Astrophys. J.}, {\bf 793}, 136 (2014)\\
26. Udalski A. et al.,  OGLE-IV: Fourth Phase of the Optical Gravitational Lensing Experiment, {\it Acta Astronom.}, {\bf 65}, 1-38 (2015)\\
27. Hynes R.I., et al., Correlated X-Ray and Optical Variability in V404 Cygni in Quiescence,  {\it Astrophys. J. Let.}, {\bf 611} L125-128 (2004)\\

\end{document}